\documentclass[journal=jacs, manuscript=letter]{achemso}
\usepackage[version=3]{mhchem} 
\SectionNumbersOn

\usepackage[utf8]{inputenc}
\usepackage[T1]{fontenc}
\usepackage{graphicx}
\usepackage{amsmath}
\usepackage{amssymb}
\usepackage{amsfonts}
\usepackage{soul}
\usepackage{dcolumn}
\usepackage{bm}
\usepackage{hyperref}
\usepackage[mathlines]{lineno}

\usepackage{xcolor}
\usepackage{titlecaps}
\definecolor{sph}{rgb}{0.0588, 0.3216, 0.7294} 
\definecolor{ppk}{rgb}{1.0, 0.4549, 0.0902} 
\newcommand{\nc}{\newcommand}
\nc{\nn}{\nonumber}
\nc{\txt}{\textrm}
\nc{\txtsup}{\textsuperscript}
\nc{\txtsub}{\textsubscript}
\nc{\calL}{\mathcal{L}}
\nc{\U}{\mathcal{U}}
\nc{\T}{\mathcal{T}}
\nc{\E}{\mathcal{E}}
\nc{\calH}{\mathcal{H}}
\nc{\calD}{\mathcal{D}}
\nc{\calJ}{\mathcal{J}}
\nc{\tilS}{\tilde{S}}
\nc{\sect}[1]{{\sl #1 .--}}
\nc{\subhajit}[1]{\textcolor{black}{#1}}
\nc{\change}[1]{\textcolor{black}{#1}}
\nc{\YD}[1]{\textcolor{black}{#1}}
\nc{\oli}[1]{\textcolor{black}{#1}}
\newcommand{\onlinecite}[1]{\hspace{-1 ex} \nocite{#1}\citenum{#1}} 
\newcommand{\ket}[1]{\left|#1\right\rangle}
\newcommand{\bra}[1]{\left\langle #1\right|}

\newcommand{\NMI}{\text{NMI}}

\author{Subhajit Sarkar}
\email{sbhjt72@gmail.com}
\affiliation{Department of Physics, School of Natural Sciences, Shiv Nadar Institution of Eminence Deemed to be University, NH91, Tehsil Dadri, Gautam Buddha Nagar, Uttar Pradesh - 201314, India}

\author{Oliver L. A. Monti}
\affiliation[UoA]
{Department of Chemistry and Biochemistry, University of Arizona, 1306 E. University Blvd., Tucson, Arizona 85721, United States \\Department of Physics, University of Arizona, 1118 E. Fourth Street, Tucson, Arizona 85721, United States}

\author{Yonatan Dubi}
\affiliation{Department of Chemistry, Ben-Gurion University of the Negev, Beer Sheva 84105, Israel,
}
\altaffiliation{Ilse Katz Center for Nanoscale Science and Technology, Ben-Gurion University of the Negev, Beer Sheva 84105, Israel.
}

\date{\today}

\title{Spinterface-like mechanism of the chirality-induced spin selectivity in donor chiral-bridge acceptor complexes}
\abbreviations{CISS}

\begin{document}

\begin{abstract}
The chirality-induced spin selectivity (CISS) effect has been invoked to explain recent reports of differences in the time-resolved EPR signals between chiral and achiral molecules. However, the microscopic origin of these differences and their connection to CISS remains contested, particularly since these systems lack a metal interface.
Here we introduce an intramolecular spinterface-like mechanism that naturally arises within donor-chiral bridge-acceptor (D--$\chi$B--A) complexes and quantitatively reproduces experimentally reported observed spin polarization in time-resolved EPR studies. In our two-electron Lindblad model, the photoexcited charge-transfer electron traversing the chiral bridge exchanges with the residual donor electron, which acts as a localized magnetic moment analogous to an induced magnetic moment on an electrode surface. The resulting through-bridge charge current produces an effective solenoidal field at the donor--bridge interface, breaking spin degeneracy and directional symmetry, thus enabling spin-selective transport without invoking intrinsic spin-orbit coupling on the bridge. We show that the interplay between this current-induced field, donor thermalization (which breaks time-reversal symmetry), and bridge spin mixing yields tens-of-percent polarization over realistic experimental conditions and charge-transfer time scales, matching reported CISS signatures in triads and DNA hairpins. By explicitly resolving the dependence on solenoidal coupling strength, temperature, and spin-mixing rates, the model identifies the regime in which internal spinterfaces can generate robust CISS-like spin filtering. These findings demonstrate that CISS-like signals in isolated D--$\chi$B--A complexes are fully compatible with a spinterface mechanism, providing a unified conceptual framework for interpreting both device-based and molecule-internal CISS platforms.
\end{abstract}


\maketitle

\section{Introduction}
The chirality-induced spin-selectivity (CISS) refers to the intrinsic preference of chiral molecules to transmit carriers with a particular spin (majority carriers) in the absence of an externally applied magnetic field, with a spin orientation that reverses with the chirality of the molecule \cite{Ray99, bloom_chiral_2024, aiello_chirality-based_2022}. CISS is positioned both as a fundamental problem - its microscopic origin and even its defining signatures are still debated \cite{evers_theory_2022, Sarkar2025SpinterfaceCISS} -- and as a technologically relevant platform for applications spanning spintronics, catalysis, energy conversion, and quantum information \cite{bloom2024chemical, aiello_chirality-based_2022, chiesa2023chirality, bloom_chiral_2024}, where robust spin selectivity over the cryogenic-to-ambient temperature regime could enable molecule-scale spin filters.

In platforms such as photoemission and transport \cite{bloom_chiral_2024}, the customary ``smoking guns'' for the CISS effect are a photoelectron spin polarization that switches with enantiomer, and a current-voltage curve that depends on the magnetic electrode magnetization and reverses when the chirality of the molecular dipole is flipped \cite{nguyen2024mechanism, aragones2025dipole}, respectively. Because these signals are tightly connected with the experimental platform, it is hard to disentangle the platform contingencies from the intrinsic role of molecular chirality. Specifically, it has been suggested that for the CISS effect to take place, the chiral molecule must be in contact with a surface, which is always the case in these platforms \cite{alwan2021spinterface, dubi2022spinterface, Sarkar2025SpinterfaceCISS}.  

The search for a new platform to observe the CISS effect, in which the molecule is isolated from contact with a surface, motivated recent CISS experiments with donor-chiral bridge-acceptor (D--$\chi B$--A) triads. Time-resolved electron paramagnetic resonance (trEPR) measurements reported that photogenerated radical pairs in such triads exhibit fingerprints of CISS, e.g. line-shape changes and significant triplet character in D--$\chi B$--A systems (compared to their achiral counterparts), and even reported $\sim$38\% CISS contribution to spin dynamics in randomly oriented organic D--$\chi B$--A enantiomers; direct CISS control of spin dynamics has also been claimed to be observed in isolated D--$\chi B$--A molecules \cite{eckvahl_direct_2023, latawiec_detecting_2025, eckvahl_detecting_2024,eckvahl2026chirality}. Across various setups, these experiments reported polarization levels ranging from $\sim 20\%$ to $\sim 60\%$ in the temperature range of $\sim 85$ to $\sim 100$~K, as determined by fitting to a model with a CISS parameter \cite{eckvahl_direct_2023, latawiec_detecting_2025, eckvahl_detecting_2024}. 

These studies are significant because they probe CISS without metal (and/or semi-metal) substrates, helping to disentangle intrinsic chiral-bridge effects from interface contributions that complicate the interface-based transport experiments \cite{ saxena2025probing, subotnik2023chiral}. At the same time, direct detection remains challenging. Some D--$\chi B$--A architectures, e.g., QD-peptide-C$_{60}$, yield spin-polarized charge-transfer states compatible with CISS but not uniquely attributable to it \cite{privitera_direct_2022, privitera_challenges_2023}. The fact that the experimental results show exactly the same signal for both enantiomers (due to the nature of the measurement) implies that while CISS can be an interpretation of the results, other effects (such as magneto-chiral anisotropy \cite{rikken2023comparing}) may also play a role. Consequently, while spin polarization observed in D--$\chi B$--A is widely attributed to CISS, its microscopic origin, ranging from weak intrinsic spin-orbit coupling augmented by electron-vibration or electron-electron interactions to possible surface-assisted mechanisms, remains an open question \cite{chiesa2024many, zhang_dynamical_2025, tiwari_role_2022, latawiec_detecting_2025, eckvahl_detecting_2024, theiler2025nonhermitian, osti_beratan_pedagogical}.

Here we propose a spinterface-like model for the possible spin polarization observed in D--$\chi B$--A complexes. In the spinterface pictures, spin-orbit and exchange interactions between spins in the molecule and in the electrode (i.e., across the interface) transduce chirality-driven orbital motion into spin, while dissipation provides time-reversal symmetry breaking \cite{alwan2021spinterface, dubi2022spinterface, monti2024surface, Sarkar2025SpinterfaceCISS}. In D--$\chi B$--A complexes a spinterface-like situation can naturally arise from within: during photoinduced charge--transfer, the {remaining} donor electron behaves as a localized magnetic moment that magnetically exchanges (i.e. interacts through magnetic exchange) with the itinerant excited electron in the chiral bridge, thereby playing the role of the ``surface moment'' and furnishing the same symmetry-breaking and amplification pathway ordinarily attributed to an interface with an electrode in transport and photoemission CISS experiments \cite{Sarkar2025SpinterfaceCISS}. The current through the bridge creates an effective solenoidal field at the donor site, removing the spin degeneracy between the up and down spins at the first bridge site, while dissipation (in our model due to thermalization in the donor and spin-mixing in the donor--bridge interface) supplies the needed time-reversal symmetry breaking. Our results demonstrate that observation of CISS in D--$\chi B$--A complexes is fully compatible with the spinterface mechanism.

\change{Unlike the existing radical pair mechanism, in which singlet–triplet interconversion is driven by an external magnetic field acting on two already-separated radical spins \cite{fay_origin_2021, luo_chiral-induced_2021, fay_enantioselective_2025, fay_radical_2019, fay_chirality-induced_2021, fay_electron_2019}, and unlike earlier Lindblad descriptions of DBA charge transport that do not resolve spin \cite{berlin_charge_2008, segal_steady-state_2001}, our spinterface model retains the chiral bridge as a spatially resolved tight-binding chain and generates spin selectivity through a self-consistent current-induced solenoidal field at the donor–bridge interface -- requiring neither intrinsic spin–orbit coupling on the bridge nor an external magnetic field. When there is no through--bridge current, our framework reduces to a bridge-resolved analogue of the radical pair description at zero field, confirming that the spinterface mechanism is the sole origin of the predicted polarization.}


\section{Model}

\begin{figure}
    \centering
    \includegraphics[width=0.55\linewidth]{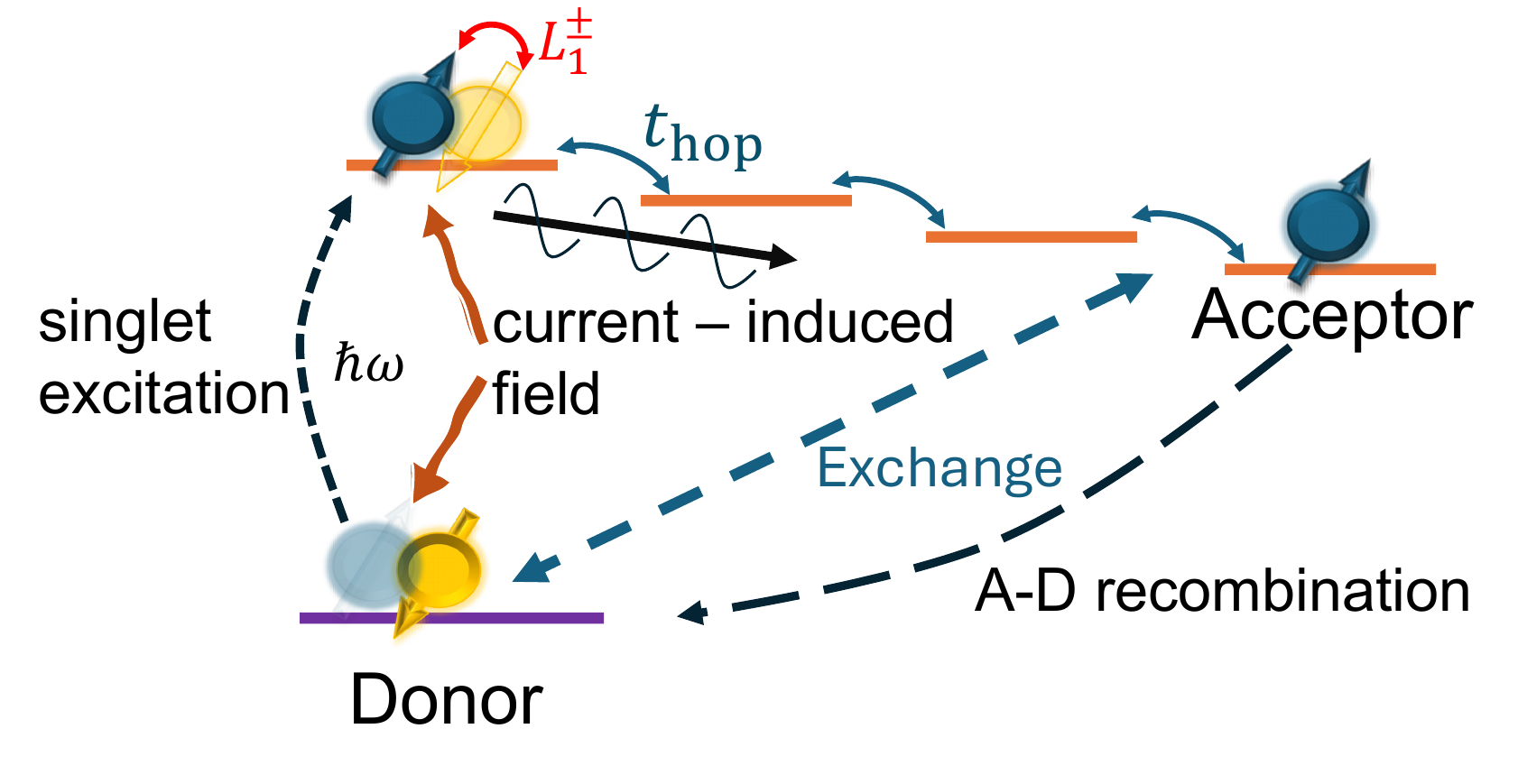}
    \caption{\textbf{Schematic depiction for the Donor-chiral-bridge-acceptor (D--$\chi$B--A) system within the Hubbard picture.}  Photoexcitation of the donor (D) injects an electron into the chiral bridge $\chi$B (with $N=4$), which propagates toward the acceptor-sink (A) with hopping amplitude $t_{\mathrm{hop}}$ and can recombine back to D. Injection and recombination are described by the Lindblad operators $L_{\mathrm{inj}}$ and $L_{\mathrm{rec}}$, while local spin mixing at the donor-bridge interface is captured by $L_{1}^{\pm}$ \change{(highlighted by the faded yellow spin to denote the spin-flip operator, rather than double occupancy of the donor excited state)}. The through-bridge current generates an effective interfacial field $g\mu_B B_{\mathrm{eff}}=\alpha_0 J$ (a ``spinterface-like" mechanism). }
    \label{fig:schematic_hubbard}
\end{figure}

\subsection{System}\label{sec:system}
We consider the D--$\chi B$--A complex to contain a maximum of two {relevant} electrons, see Fig. \ref{fig:schematic_hubbard}, the closest scenario to that of Ref.~\onlinecite{eckvahl_direct_2023}. The donor (D) hosts two electronic orbitals \( \alpha\in\{a,b\} \) that are otherwise degenerate, but the degeneracy is broken {in a Hubbard sense} by the Coulomb and spin-exchange interactions between the donor electrons. 

The chiral bridge ($\chi B$) is a one-dimensional chain of \(N\) sites \( j=1,\dots,N \), and the acceptor (A) is treated implicitly as a dissipative sink attached to the terminal bridge site, Fig. \ref{fig:schematic_hubbard}. 
The effective Hamiltonian for the donor is therefore \(\displaystyle \calH_{\text{donor}} =\sum_{\substack{\alpha = a,b, \ \sigma = \uparrow, \downarrow}}\varepsilon_{\alpha, \sigma}\, n_{\alpha\sigma}\), where $\varepsilon_{\alpha, \sigma}$ corresponds to the non-degenerate energy of the individual spin-dependent donor states. 

The bridge is described as a spin- and charge-conserving tight-binding chain with site energies \(\{\varepsilon_j\}\) and nearest-neighbor hopping \(t_{\text{hop}}\), 
\begin{equation}
\calH_{\text{bridge}}=\sum_{j,\sigma}\varepsilon_j\, n_{j\sigma} -t_{\text{hop}}\sum_{j=1}^{N-1}\sum_{\sigma}\!\Big(c^\dagger_{j+1,\sigma}c_{j\sigma}+\text{h.c.}\Big)~.
\end{equation} Note that the bridge does not contain any explicit spin-orbit coupling, unlike most existing models of CISS \cite{theiler2025nonhermitian, chiesa2024many, zhang_dynamical_2025}. {Bridge chirality will enter through the generation of a magnetic field, parallel to the molecular chiral axis, generated when current passes through it, as described in what follows}. 
\par
Exchange interactions between the spins in the donor and the bridge are described by an isotropic (Heisenberg) exchange term,  
\begin{equation} \displaystyle \calH_{\text{ex}}^{(DB)} =  \!\sum_{j=1}^{N} J_{j} \mathbf{S}_D \cdot \mathbf{S}_j~. 
\end{equation}
This can arise from interfacial superexchange and transfers the donor’s spin information to the itinerant electron on the bridge. Within the donor, the exchange \(\calH_{\text{ex}}^{(D)} = J_{D}\,\mathbf{S}_a \cdot \mathbf{S}_b\) fixes the singlet-triplet splitting, thereby determining which donor spin configuration is energetically accessible, see Fig~\ref{fig:schematic_state}(a).

Moreover, any external or internal effective magnetic field $B_{\mathrm{eff}}$ acts as a field bias parameter,
\(\calH_Z = - \mu_B B_{\mathrm{eff}}
\left(g_D S_{D}^{z} + g_B S_1^{z}\right).\)
The central premise of our model is that the Zeeman term acts at the interface, i.e., on both the donor and on the first bridge site. This is equivalent to singlet-triplet splitting of the donor-bridge (DB) state, see Fig.~\ref{fig:schematic_state}(a). The first bridge site corresponds to an effective description of {the initially excited charge--transfer state} in the molecule. The effective magnetic field splits the triplet states in the donor as well as the triplet DB states. The splitting at the first bridge site is determined only by the electron spin on that site. The term \(\calH_{\text{ex}}^{(D)}\) does not modify this splitting; it only correlates the donor and bridge spins by shifting the relative energies and thus populations of the coupled donor–bridge spin configurations. 

Thus, the system can be minimally described by the following Hamiltonian:
\(
\calH \;=\; \calH_{\text{donor}} \;+\; \calH_{\text{bridge}} \;+\; \calH_{\text{Z}} \;+\; \calH_{\text{ex}}^{(D)} \;+\; \calH_{\text{ex}}^{(DB)} .
\) 
We consider the system through its density matrix $\rho$, which follows the Lindblad master equation \cite{lindblad1976generators,gorini1976completely,manzano2020short}
\( \displaystyle
\dot{\rho}
= -\,i[\calH,\rho]
+ \sum_k \Big(L_k \rho L_k^\dagger - \tfrac12\{L_k^\dagger L_k,\rho\}\Big).
\)
At most one itinerant electron resides on the bridge during the charge-transfer (CT) process, and the remaining electron stays on the donor, effectively acting as a localized moment. Charge transfers and other environmental effects are encoded by so-called jump operators \( \{L_k\} \). In our model, (i) photoexcitation and recombination occur incoherently  (i.e., classical photons) described by \(L_{\rm inj}\) and \(L_{\rm rec}\) operators \cite{sarkar2020environment}; (ii) the bridge hosts at most one itinerant electron; and (iii) the acceptor is treated implicitly as a right-end sink. 

\subsection{Charge-transfer processes}\label{sec:CT}

\begin{figure}
    \centering
    \includegraphics[width=0.65\linewidth]{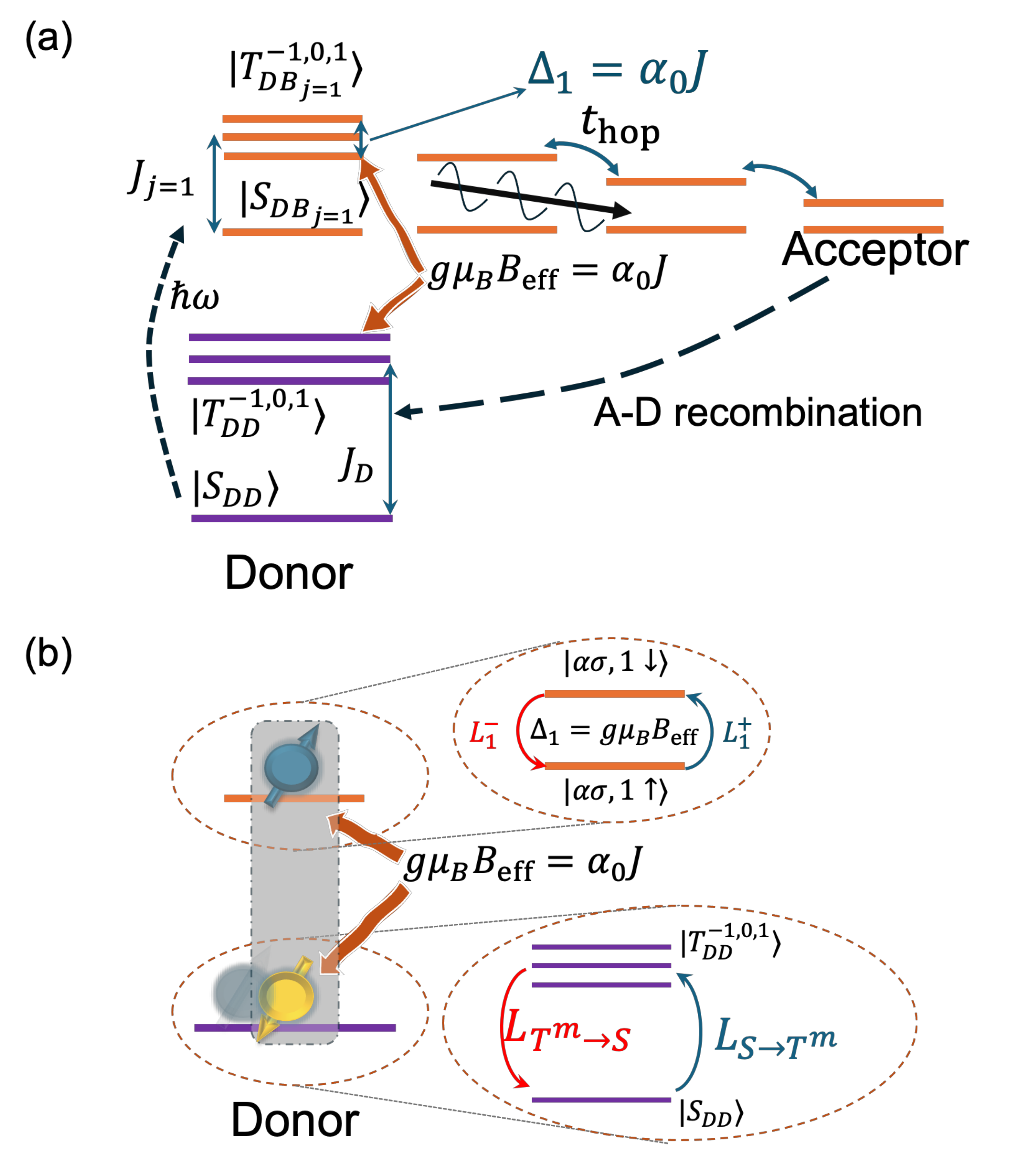}
    \caption{\textbf{(a)} \textbf{Schematic of D--$\chi$B--A complex in terms of its singlet-triplet  state manifold}. The bare singlet-triplet gap in the donor is governed by the exchange coupling $ J_D$. The same exchange coupling in the donor-bridge manifold is distance-dependent, with $J_j$ being smaller for transfer to a more distant bridge site.
    \textbf{(b)} \textbf{Thermalization mechanism at the interface.} 
    The current-induced field produces a Zeeman splitting $\Delta_1=g\mu_B B_{\mathrm{eff}}$ of the single-electron levels on the first bridge site, enabling local spin-flip jumps $L_{j = 1}$ and $L_{j = 1}^\dagger$ between \(|\alpha\sigma,1\uparrow\rangle\) and \(|\alpha\sigma,1\downarrow\rangle\). The donor spin couples to bridge spins via exchange $J_j\,\mathbf{S}_D\!\cdot\!\mathbf{S}_j$, and donor singlet-triplet interconversion is modeled by the dissipative channels $L_{S\rightarrow T^m}$ and $L_{T^m\rightarrow S}$ between \(|S_{DD}\rangle\) and \(|T_{DD}^{m}\rangle\) with \((m=-1,0,1)\).}
    \label{fig:schematic_state}
\end{figure}
At equilibrium, the donor is occupied by two electrons. The associated two-electron spin (doubly-occupied donor, DD) manifold consists of one singlet and three triplets, which we denote as $\ket{\chi_{DD}}\in\big(\ket{S_{DD}},\ket{T^{m}_{DD}}\big)$, where $m\in\{-1,0,+1\}$ labels the triplet spin projection. Explicitly, $\ket{S_{DD}}=\frac{1}{\sqrt{2}}\big(\ket{\alpha\uparrow,\alpha'\downarrow}-\ket{\alpha\downarrow,\alpha'\uparrow}\big)$, $\ket{T^{+1}_{DD}}=\ket{\alpha\uparrow,\alpha'\uparrow}$, $\ket{T^{0}_{DD}}=\frac{1}{\sqrt{2}}\big(\ket{\alpha\uparrow,\alpha'\downarrow}+\ket{\alpha\downarrow,\alpha'\uparrow}\big)$, and $\ket{T^{-1}_{DD}}=\ket{\alpha\downarrow,\alpha'\downarrow}$. {Photoexcitation} transfers one electron from the donor to the first bridge site while the other electron remains on the donor, thereby creating a donor-bridge ($DB$) configuration. The only optically allowed transition is from the singlet donor ground state to a singlet donor-bridge excited state, described by the Lindblad operator $L_{\mathrm{inj}}=\sqrt{\eta}\,\ket{S_{DB}(1)}\bra{S_{DD}}$, where $\eta$ is the rate of transition, as indicated in Fig~\ref{fig:schematic_state}. Immediately after injection, the electron resides on the leftmost bridge site ($j=1$ in Fig.~\ref{fig:schematic_hubbard}); subsequent coherent hopping moves it to other bridge sites \cite{iv2020ballistic}. To track this, we introduce the site-resolved $DB$ spin manifold $\ket{\chi_{DB}(j)}\in\big(\ket{S_{DB}(j)},\ket{T^{m}_{DB}(j)}\big)$ with $j=1,\dots,N$, where $\ket{S_{DB}(j)}=\frac{1}{\sqrt{2}}\big(\ket{\alpha\uparrow,j\downarrow}-\ket{\alpha\downarrow,j\uparrow}\big)$, $\ket{T^{+1}_{DB}(j)}=\ket{\alpha\uparrow,j\uparrow}$, $\ket{T^{0}_{DB}(j)}=\frac{1}{\sqrt{2}}\big(\ket{\alpha\uparrow,j\downarrow}+\ket{\alpha\downarrow,j\uparrow}\big)$, and $\ket{T^{-1}_{DB}(j)}=\ket{\alpha\downarrow,j\downarrow}$. {Without loss of generality}, the first bridge site ($j=1$) is identified with the donor-excited state. Importantly, triplet population is not created directly by optical excitation; instead, it emerges downstream during the chiral-bridge charge-transfer cycle via through-bridge spin mixing, followed by spin-conserving recombination and subsequent donor thermalization, as described below.

The terminal bridge site ($j=N$) acts as an acceptor-like sink. Recombination returns the charge-transferred $DB$ configuration back to the donor $DD$ manifold without changing the two-electron spin state: the electron tunnels back to the donor in a spin-conserving way. This means that the singlet channel $\ket{S_{DB}(N)}$ recombines only into $\ket{S_{DD}}$, while each triplet component $\ket{T^{m}_{DB}(N)}$ recombines into the corresponding donor triplet $\ket{T^{m}_{DD}}$ with the same $m\in\{-1,0,+1\}$. We encode this as spin-resolved Lindblad jumps with recombination rate $\gamma$, namely 
$L_{\mathrm{rec}}=\lbrace \sqrt{\gamma}\,\ket{S_{DD}}\bra{S_{DB}(N)}; \sqrt{\gamma}\,\ket{T^{m}_{DD}}\bra{T^{m}_{DB}(N)} \rbrace$, corresponding to the singlet and the triplet ($m = \lbrace -1,0,+1 \rbrace$) states, respectively, so that recombination is explicitly spin--conserving. This closes the donor $\to$ bridge $\to$ acceptor cycle within the reduced $DB$ subspace.

Between 
{photoexcitation} and recombination, the donor-bridge charge-transfer state is not spin-isolated: the bridge spin is coupled to a thermal environment that drives incoherent spin flips, and this dissipation provides the mechanism that converts optically prepared singlet character into triplet character {or vice versa} in a controlled, thermodynamically consistent way by satisfying the detailed balance criterion. We model this with local Lindblad jump operators $L_{j}^{+}=\sqrt{\gamma_{j}^{d\to u}}\,S_{j}^{+}$ and $L_{j}^{-}=\sqrt{\gamma_{j}^{u\to d}}\,S_{j}^{-}$, where the flip bias is set by the local Zeeman splitting $\Delta_{j}\equiv E_{\downarrow}-E_{\uparrow}=\mu_{B}g_{B,j}B_{\mathrm{eff}}$, as indicated in Fig~\ref{fig:schematic_state}(b). Thermodynamic consistency is enforced by imposing detailed balance directly at the level of the rate ratio, $\gamma_{j}^{u\to d}/\gamma_{j}^{d\to u}=e^{-\beta\Delta_{j}}$. A convenient parametrization is $\gamma_{j}^{d\to u}=\gamma_{m}\,r_{j}$ and $\gamma_{j}^{u\to d}=\gamma_{m}\,r_{j}\,e^{-\beta\Delta_{j}}$, with $r_{j}\ge 0$ a site-dependent prefactor and $\gamma_{m}$ setting the overall \emph{spin--mixing} rate. In our model, this channel is interface-local: We take $r_{1}=1$ and $r_{j>1}=0$, so spin flips act only at the donor--bridge hybrid site $j=1$ where the effective splitting is operative, while downstream bridge sites are spin-unresolved in this respect. Microscopically, such interface-local mixing can arise from transverse magnetic noise due to hyperfine fields on nearby nuclei and/or vibronic modulation of spin-orbit and $g$-tensor anisotropies at the donor-bridge junction. Furthermore, reversing $B_{\mathrm{eff}}$ flips the sign of $\Delta_{1}$ and therefore reverses the preferred direction of $S^{+}_{j}$ versus $S^{-}_{j}$ jumps, selecting a different favored triplet projection. In this way, triplet population is generated during charge transfer: bridge spin mixing converts singlet-born $DB$ states into configurations with finite triplet weight, which then returns to the donor under spin-conserving recombination.

After recombination, the spin-correlated pair evolves within the donor manifold. The donor exchange Hamiltonian $\calH_{\mathrm{ex}}^{(D)}$ sets a large singlet-triplet gap $\Delta_{ST}$ (in the singly occupied $a$-$b$ sector this splitting is controlled by the exchange coupling $J$ and can be as large as $\mathcal{O}(1~\mathrm{eV})$), so coherent singlet-triplet mixing inside the donor is strongly suppressed; by contrast, the triplet sublevels remain near-degenerate and can be weakly admixed among themselves by hyperfine and other small anisotropic interactions. In parallel, we include a fast, featureless electronic environment, e.g., from the embedding medium, with an approximately flat coupling scale $\gamma_{0}$ that thermalizes donor populations. We model intradonor thermalization using spin-selective Lindblad jumps $L_{T^m\to S}=\sqrt{\Gamma_{T\to S}}\,\ket{S_{DD}}\bra{T^{m}_{DD}}$ with $m\in\{-1,0,+1\}$, which relax donor-triplet population into the donor singlet, and we parameterize $\Gamma_{T\to S}=\gamma_{0}$. To maintain thermodynamic consistency, we optionally include the reverse, thermally activated backflow $L_{S\to T^m}=\sqrt{\Gamma_{T\to S}e^{-\beta\Delta_{ST}}}\,\ket{T^{m}_{DD}}\bra{S_{DD}}$, so that the rate ratio satisfies detailed balance, $\Gamma_{S\to T}/\Gamma_{T\to S}=e^{-\beta\Delta_{ST}}$, see Fig~\ref{fig:schematic_state}(b). Consequently, in the absence of driving, the donor subspace relaxes toward the Boltzmann distribution at inverse temperature $\beta$, with the singlet favored when $\Delta_{ST}>0$. Within the experimental photochemical cycle, the transient triplet population generated during charge transfer is therefore returned toward the donor singlet on longer timescales by this detailed-balance-preserving thermalization channel.

\subsection{The Spinterface mechanism}
The spinterface mechanism of CISS for the D--$\chi B$--A complexes is encoded by two components. The first is the interface exchange interactions (described by $\calH_{\text{ex}}^{(DB)}$ above). The second is the effective magnetic field due to the charge transfer process along the chiral bridge (i.e., a solenoid field) that is felt by the localized spins at the donor, {both in the ground and excited states}. 

The charge transfer through the $\chi B$ is quantified by the spin-resolved bond current operator, \( \hat{J}_{i,\sigma} = i\,t_{\text{hop}} ~\left(c^\dagger_{i+1,\sigma}c_{i\sigma}-c^\dagger_{i,\sigma}c_{i+1,\sigma}\right) \). 
In the steady state, due to charge conservation, the charge-current through all bonds is the same, and thus \(\displaystyle \hat{J} = \sum_{\sigma} \hat{J}_{i,\sigma}\). {Here the molecular chirality enters, through the generation of a} current-induced effective field, taken to be aligned with the molecular chiral axis, \( \displaystyle g\mu_B{\bf B}_{\text{eff}}=\alpha_0\,(J_{\uparrow}+J_{\downarrow})\,\hat{\bf z} \) \cite{monti2024surface}, where \( \displaystyle J_{\sigma}=\mathrm{Tr}\!\left[\hat{J}_{\sigma}\rho_{\mathrm{SS}}\right] \) is the steady-state spin-resolved current, \(\rho_{\mathrm{SS}}\) the steady-state density matrix, and \(\alpha_0\) (in \(\mu\)eV/nA) parametrizes the solenoidal-field strength, including the geometric Oersted field of the helical current, interactions and vibronic enhancements \cite{medina2015continuum,salazar2018spin,ludena2025toward,fransson2020vibrational}. \subhajit{We note that an interface must be localized in space, a requirement that leads us to identify the interface as the donor and the first bridge sites. Therefore, the interfacial field acts on these two}
sites, giving \( \displaystyle \mathcal{H}_{\text{Z}}=-\,g\mu_B{\bf B}_{\text{eff}}\!\cdot\!({\bf S}_D+{\bf S}_1)=-\,\alpha_0\,(J_{\uparrow}+J_{\downarrow})\,(S_D^{z}+S_1^{z}) \), which lifts the spin degeneracy locally at the donor-bridge interface and biases the ensuing spin dynamics and steady-state spin polarization. Because \(\mathcal{H}_{\text{Z}}\) depends on \(J_{\uparrow,(\downarrow)}\), \(\rho_{\mathrm{SS}}\) and \(J_{\uparrow,(\downarrow)}\) must be determined self-consistently.


The spinterface approach to the CISS effect in molecular junctions highlighted an amplification mechanism for the solenoid field through exchange interactions \cite{alwan2021spinterface, dubi2022spinterface, Sarkar2025SpinterfaceCISS}. It is useful to describe the mechanism that leads to spin-polarization in our system in terms of the same ``amplification mechanism'',  corroborated by numerical results in the following sections. One way to think about the process is as follows: After the initial excitation, there is complete spin-degeneracy up and down spins in the bridge. As electrons flow through the bridge, a ``solenoid field'' is generated, which breaks the symmetry at the donor site, which is now slightly more amenable to excitations of a given spin. This leads to a spin imbalance on the bridge, which, through exchange interactions {and the thermalization within the donor levels}, enhances the spin asymmetry at the donor site. It is now even more favorable to excite majority spins from the donor, which enhances the spin imbalance at the bridge even more, and so on. Eventually, the steady state has a broken spin-degeneracy, leading to a majority-spin being injected from the donor to the bridge and to an observed spin-polarization. 



\section{Results}


We define the CISS polarization, $P$, as the ratio of spin and charge densities, $\mathrm{P}( {\rm in } \%) \equiv 100\times \frac{2\langle S^z_N\rangle}{\langle n_N\rangle}$, where the spin density $\langle S^z_N\rangle $ and the charge density $\langle n_N\rangle$ are evaluated in the steady state, and $\langle \hat{O} \rangle = {\rm Tr}[\rho_{SS} \hat{O}]$. When comparing with the experiment, we compare $P$ with the phenomenological `CISS fraction' parameter used in trEPR fits, i.e., as a measure of the initial spin imbalance of the radical-pair manifold. In what follows, we plot the polarization $P$ versus the magnitude of the solenoidal coupling $\alpha_0\!\ge\!0$ expressed in units of eV/nA. The direction of charge transfer sets the sign of the effective field, $\mathbf B_{\rm eff}=\alpha_0\,|J|\,{\rm sgn}(J)\,\hat{\mathbf z}$, so reversing the current (${\rm sgn}(J)\!\to\!-{\rm sgn}(J)$) flips the {sign of the spin of the} donor ground state and hence the sign of $P$, and the total current is defined as $J = J_{\uparrow} + J_{\downarrow}$. This is equivalent to switching the enantiomer, which maps $\hat{\mathbf z}\!\to\!-\hat{\mathbf z}$ at a fixed current. In the following calculation, we take $\gamma = \eta = \gamma_0 = t_{\rm hop}$, although this choice does not qualitatively change the results.

{As we show in what follows,} our model reproduces the correct order of magnitude of the polarization. For realistic currents inferred from the reported through-bridge charge-transfer times, our $P$ values reach a few tens of percent, comparable to the ``apparent CISS fraction'' extracted from trEPR: $38 \pm 4$\% at X-band for the organic triads, and between $21 \pm 5$\% and $29 \pm 8$\% at Q/W-band, respectively, while DNA hairpins acquire 29\% at X-band and 62\% at Q-band in representative fits \cite{eckvahl_detecting_2024, eckvahl_direct_2023,latawiec_detecting_2025}. The temperature dependence of $P$ trends maintains this magnitude within the $85-105$ K window used in the experiments, with higher temperatures reducing polarization as the thermal population redistributes donor levels. {This suggests that the spinterface model developed here is capable of generating the observed experimental spin signatures, and we next investigate the key factors that contribute to the ``CISS fraction''.}

\subsection{Role of the solenoid field}
\begin{figure}
    \centering
    \includegraphics[width=0.7\linewidth]{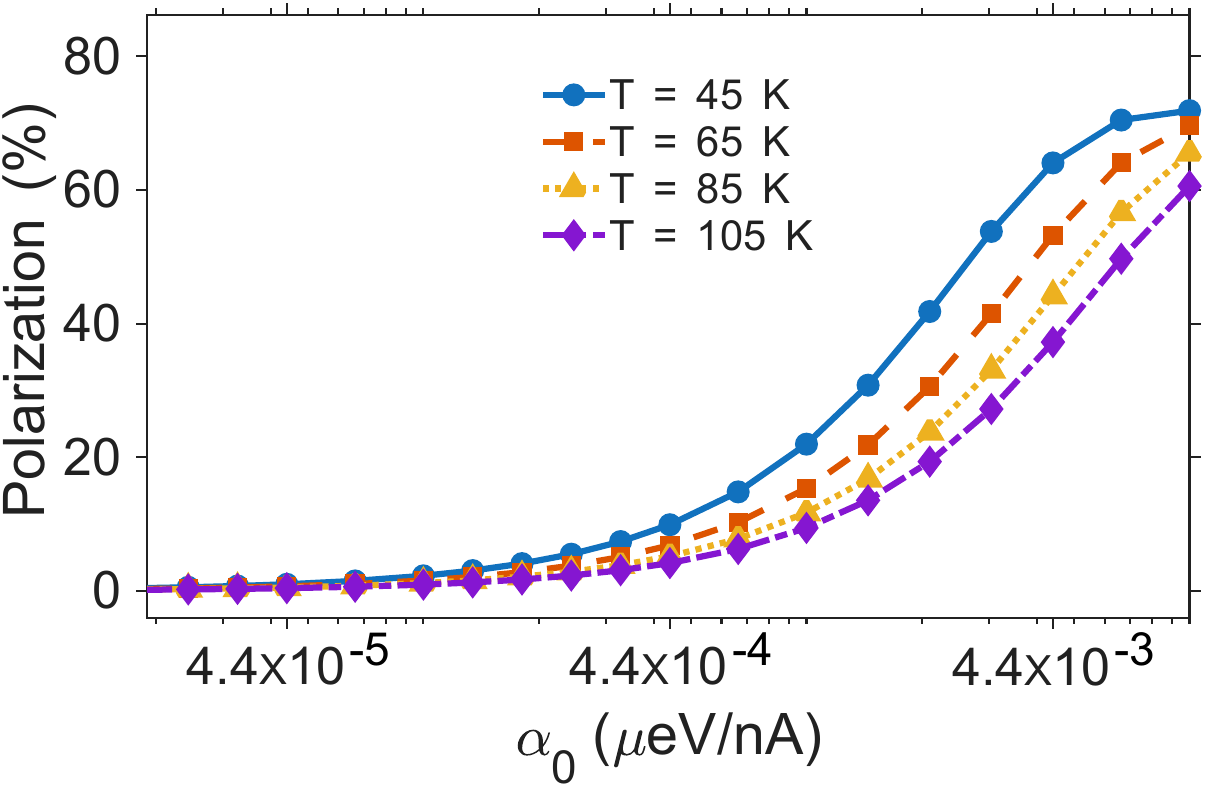}
    \caption{\textbf{Spin--polarization vs. effective strength of the solenoid field at different temperatures:} Plot of the steady-state spin--polarization $P$ versus the solenoidal-coupling magnitude $\alpha_0$ on a log scale at a fixed mixing rate $\gamma_m = 0.014 t_{\rm hop}$; curves correspond to four temperatures (see legend).}
    \label{fig:pol_vs_alpha0}
\end{figure}
Figure~\ref{fig:pol_vs_alpha0} shows the steady-state polarization $P$ as a function of the magnitude of the solenoidal coupling $\alpha_0$ for a fixed value of $\gamma_m = 0.014t_{\rm hop}$. 
In the small-$\alpha_0$ regime, $B_{\rm eff} \propto \alpha_0 J$ weakly biases the donor singlet-triplet manifold and $P$ grows with $\alpha_0$. As $\alpha_0$ increases further, the bias competes with the spin-mixing effects in the bridge, and thus polarization $P(\alpha_0)$ crosses over to a slower and saturating growth, set by $\gamma_m$ and $T$. Lowering $T$ increases the polarization. The four temperatures used here ($45 - 105$ K) lie squarely within the experimental trEPR range: fully organic triads and DNA hairpins are measured at 85 K \cite{eckvahl_direct_2023, eckvahl_detecting_2024}, while QD–peptide–C$_{60}$ is probed at $40$ K (with additional datasets at $10$ K) \cite{privitera_direct_2022}. 

\subsection{Magnitude and interpretation of \texorpdfstring{$\alpha_0$}{alpha0}}
A purely geometric estimate for a helical bridge with DNA-like value of pitch length $p\!\sim\!3.4$~nm gives $B_{\rm sol}/J=\mu_0/p\approx 0.37~\mu$T/nA, \cite{di2011dna,Sarkar2025SpinterfaceCISS} corresponding to a  solenoidal coupling $\alpha_{\rm geom}\sim 4.4\times10^{-5}~\mu\text{eV/nA}$.
To keep assumptions transparent in our model, we explicitly separate three tiers: (i) the strictly Oersted (purely geometric) baseline $\alpha_{\rm geom}$ estimated above; (ii) the working coupling $\alpha_0$ used in all plots; and (iii) a broad upper envelope $\alpha_{\max}$ that could arise if additional physics enhances the local effective field. Within our context, the $0.37~\mu$T/nA baseline corresponds to the DNA-hairpin platform \cite{eckvahl_detecting_2024}. \change{To rigorously ground our effective solenoidal field parameter in atomic reality, we mapped the charge-transfer path onto the exact molecular geometry of the $\NMI_2$ bridge, which possesses a discrete ground-state dihedral twist of approximately $\Theta = 100^\circ$ \cite{phan2025ab, eckvahl_direct_2023}. Integrating the discrete bond current over this fractional helical arc via the classical Biot–Savart law yields a geometric baseline of $\alpha_{\text{geom}} \approx 5.2 \times 10^{-5}\;\mu\text{eV/nA}$. This atomistically estimated value aligns remarkably well with our initial macroscopic estimate based on a continuous DNA-like helix ($\approx 4.4 \times 10^{-5}\;\mu\text{eV/nA}$), confirming that the magnitude of the spinterface field is a highly robust consequence of the triad's physical structure rather than a parameterized assumption (see Section S1 for the full derivation), and establishing the generality of our model.}

Biological radical-pair spin dynamics are typically governed by localized hyperfine and dipolar fields of $\sim 0.1$--$5$~mT \cite{efimova2008role, wong2023magnetic}. For the CISS--effect to override this background, the effective magnetic field ($B_{\mathrm{eff}}$) must reach a commensurate scale of $\sim 1$~mT, which we take as an upper-bound. For a charge-transfer current of $J \sim 1$~nA inferred from the experimental charge transfer time scale $\tau_{\rm CT}$ ($\sim 200$~ps, yielding $J \sim 0.2$--$1$~nA), achieving this field requires a Zeeman energy ($\alpha_0 J$) comparable to the electron magnetic moment ($g\mu_B \approx 0.115~\mu$eV/mT). \change{Moreover, it is critical to distinguish this primary, geometric origin of CISS from the secondary, modulatory effects of molecular vibrations. The low-frequency torsional phonons act as a mechanism of dynamic structural chirality for this non-helical molecule by rendering the charge hopping $t_{\rm hop}$ strictly dependent on the vibrationally modulated twist angle, thereby increasing $B_{\rm eff}$ non-linearly with $\Theta$, see Section S1. Unlike the existing phonon-induced mechanism of CISS, where, without the explicit spin--phonon coupling, the spin-symmetry breaking mechanism itself disappears \cite{fransson2020vibrational, fransson2021charge, fransson2022chiral, fransson2023chiral}, our mechanism does not require a spin-phonon coupling.}

These facts physically motivate an upper envelope for the effective coupling strength: $\alpha_{\max} \sim g\mu_B (B_{\mathrm{eff}}/J) \lesssim 0.1~\mu\mathrm{eV/nA}$.
Although this envelope far exceeds the bare classical solenoidal field ($\sim 0.37~\mu$T), such amplification is expected from the strong interplay of molecular curvature, exchange interactions, and vibrational effects inherent to chiral environments \cite{medina2015continuum,salazar2018spin,ludena2025toward,fransson2020vibrational}.
Crucially, however, the coupling values actually employed in this work are much smaller: we take $\alpha_0 \approx 4.4\times10^{-5} - 4.4\times10^{-3} ~\mu\text{eV/nA}$, i.e., at most a modest ($\lesssim 10^2$) multiple of the purely geometric baseline $\alpha_{\rm geom}$ and well below $\alpha_{\max}$. This choice neither invokes unphysically large internal fields nor relies on system-specific fine-tuning, but instead falls naturally within the established internal field scales of {such molecular} systems.

\subsection{Effect of spin-mixing in the chiral-bridge}

\begin{figure}
    \centering
    \includegraphics[width=0.6\linewidth]{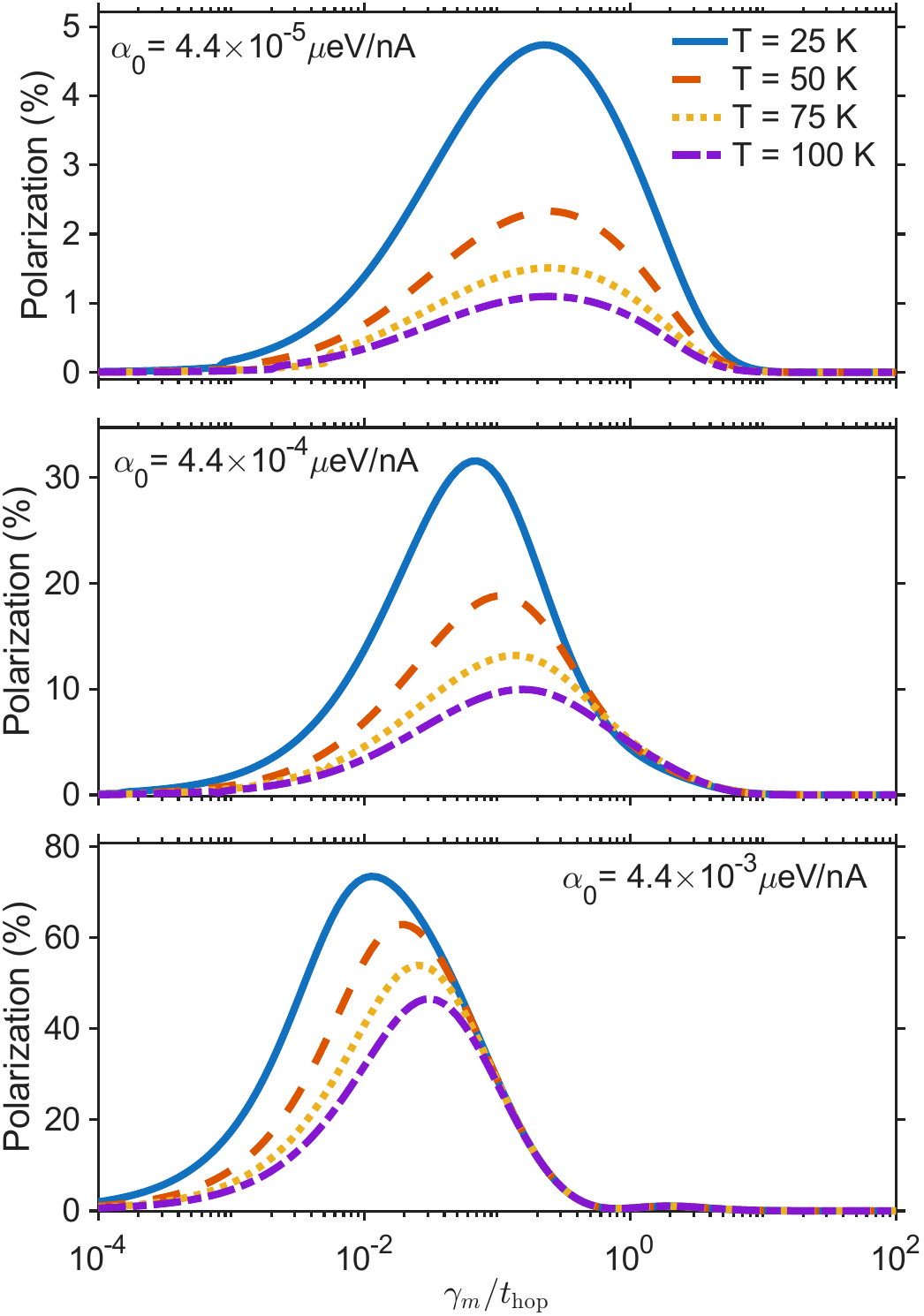}
    \caption{\textbf{Effect of spin mixing on the CISS-polarization:} Plot on a $\log-$linear of CISS-polarization $P$ in $\%$ with the relative mixing strength $\gamma_m/t_{\rm hop}$ in the first bridge site for three different effective strengths of the solenoid coupling strength $\alpha_0$, top, middle, and bottom panels, respectively, for four different temperatures.}
    \label{fig:pol_vs_mixing}
\end{figure}

The dissipative spin-mixing rate $\gamma_m$ coarse-grains chemistry-agnostic channels in D--$\chi B$--A systems, such as local magnetic noise from the surrounding chemical environment and other extrinsic fluctuations \cite{privitera_direct_2022,eckvahl_direct_2023,eckvahl_detecting_2024,latawiec_detecting_2025}, into a single phenomenological measure of the strength and timescale of incoherent spin conversion experienced by the charge-transfer state at the donor-bridge interface.

Figure~\ref{fig:pol_vs_mixing} quantifies how the bridge spin-mixing rate $\gamma_m$ competes with the current-induced donor bias $g\mu_B B_{\rm eff}=\alpha_0(J_\uparrow+J_\downarrow)$. We plot the CISS polarization $P$ (in \%) as a function of the spin-mixing strength $\gamma_m$ (in units of the bridge hopping $t_{\mathrm{hop}}$) for three solenoidal couplings $\alpha_0=4.4\times10^{-4},\,4.4\times10^{-3},\,4.4\times10^{-2}~\mu\mathrm{eV}/\mathrm{nA}$ and at four temperatures. For each $(T,\alpha_0)$, the response is distinctly nonmonotonic in $\gamma_m$, reflecting that, within our model, polarization is neither generated at injection (the optical jump prepares a singlet) nor during coherent, spin-conserving hopping along the bridge, but rather through an interface-local dissipative conversion step whose up/down rates are constrained by detailed balance. \subhajit{Most notably, our model does not incorporate an explicit SOC.} In the weak-mixing regime $\gamma_m/t_{\mathrm{hop}}\ll 1$, the photo-injected electron resides at the donor-bridge hybrid (site $j=1$) for too short a duration for the bath-induced jumps to execute an appreciable number of biased flips before the electron moves away from the interface and the cycle progresses toward recombination. Consequently, the singlet-born $DB$ state retains essentially singlet character, the transport-relevant triplet components remain weakly populated, and the terminal spin expectation $\langle S^z_N\rangle$ and hence $P$ is small. As $\gamma_m$ is increased to intermediate values, the bath has just enough time to convert a non-negligible fraction of the injected singlet content into triplets while the electron remains dynamically connected to the interface. Crucially, this conversion is not symmetric: the local Zeeman splitting $\Delta_1=\mu_B g_{B,1}B_{\rm eff}$ biases the up/down flip rates through the thermodynamically consistent ratio $\gamma^{u\to d}/\gamma^{d\to u}=e^{-\beta\Delta_1}$, so the newly generated triplet population carries a net spin projection whose sign follows the sign of $B_{\rm eff}$ and the associated direction of the net charge flow $J_\uparrow+J_\downarrow$. This spin imbalance is then transported in an essentially spin-conserving manner to the terminal site, producing a maximal polarization. In the strong-mixing regime $\gamma_m/t_{\mathrm{hop}}\gg 1$, the same dissipative channel becomes counterproductive: repeated flips dominate the interface dynamics, rapidly over-mixing the bridge spin before coherent transport can ``use'' the bias to build a directed, transport-surviving spin imbalance. In this limit, the $DB$ sector is driven toward an effectively depolarized steady state {with equipartition in the spin-manifold}, and the polarization collapses despite the abundance of flips. 

The resulting peak at intermediate $\gamma_m$ is therefore a direct fingerprint of a competition of timescales between coherent charge transfer and biased dissipative spin conversion. Quantitatively, this can be rationalized by comparing the mixing time $\tau_{\rm mix}=\hbar/\gamma_m$ with the intrinsic electronic timescale of the bridge, $\tau_{\rm hop}=\hbar/t_{\rm hop} \approx 8~\mathrm{fs}$ (using $t_{\rm hop}\approx 80~\mathrm{meV}$ \cite{phan2025ab}). Operationally, the optimal mixing rate $\gamma_m^\ast$ where $P$ is maximal lies in the range $\gamma_m^\ast/t_{\rm hop}\sim 10^{-2}$--$10^{-1}$, yielding an optimal spin-conversion window of $\tau_{\rm mix}^\ast \approx 0.08$--$0.8$~ps. Physically, this indicates that peak polarization is achieved when interface spin-flips act repeatedly during a single effective charge-transfer lifetime ($\tau_{CT} \approx 211~$ps \cite{eckvahl_detecting_2024}), yet remain slow enough ($\tau_{\rm mix} \gg \tau_{\rm hop}$) to avoid the depolarizing steady-state of the over-mixed regime. Temperature and $\alpha_0$ tune this competition in a thermodynamically consistent way: increasing $T$ weakens the detailed-balance asymmetry ($e^{-\beta\Delta_1}\to 1$), requiring faster mixing (shifting $\gamma_m^\ast$ to larger values) to achieve the same net conversion, whereas increasing $\alpha_0$ amplifies $\Delta_1$, shifting $\gamma_m^\ast$ downward while strengthening the bias and increasing the peak polarization.

We emphasize that this effective conversion window is a model-internal timescale tied to the microscopic electronic dynamics ($t_{\rm hop}$) and the cycle closure via recombination, and should not be identified with the experimentally reported bridge$\to$acceptor charge-transfer time $\tau_{\rm CT}$, which is an environment-dressed, rate-limited kinetic timescale. Crucially, these sub-ps scales corresponding to the $\tau_{\rm mix}$ are far shorter than the experimentally reported CT times of order $\sim 200~\mathrm{ps}$ \cite{eckvahl_detecting_2024,latawiec_detecting_2025}, supporting the view that the polarization is governed by fast, interface/bridge electronic dynamics rather than by the slower, environment-dressed escape step.

\subsection{Temperature dependence}

\begin{figure}
    \centering
    \includegraphics[width=0.7\linewidth]{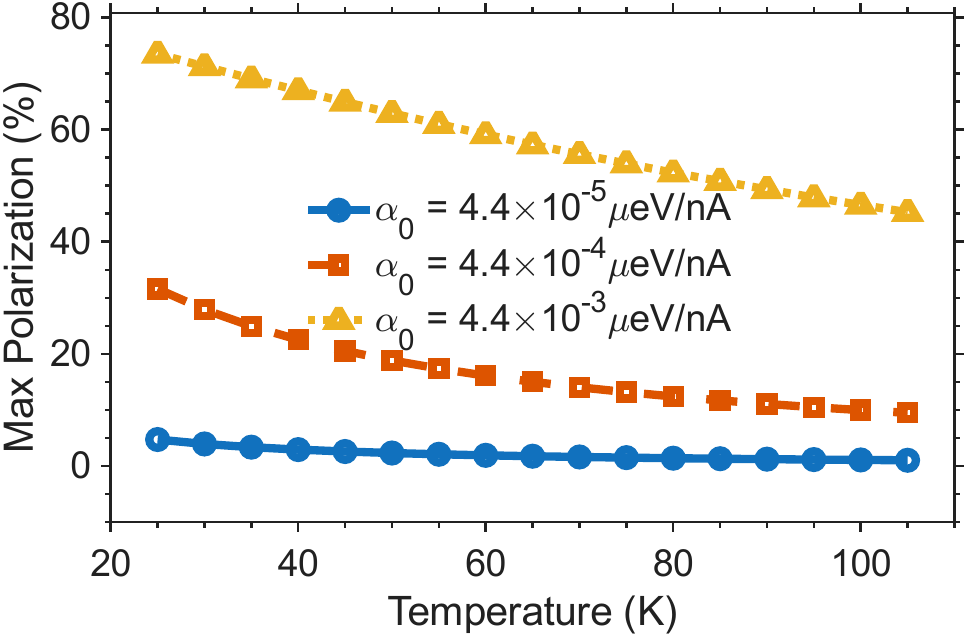}
    \caption{\textbf{Temperature dependence of CISS-polarization for different effective strengths of the solenoid fields:} Plot of the steady-state CISS polarization $P(\gamma_m^{\ast})$ with temperature $T$ for four different solenoidal-coupling magnitudes $\alpha_0$ (see legend) on a linear scale.}
    \label{fig:pol_vs_temp}
\end{figure}

Figure~\ref{fig:pol_vs_temp} displays the steady-state polarization as a function of temperature for three values of $\alpha_0$ and a fixed bridge spin-mixing rate $\gamma_m=\gamma_m^{\ast}$. In our dynamics, temperature enters through detailed-balance relaxation within the donor manifold and at the effective interface, i.e., at both the donor and the first bridge site. The monotonic decrease of the peak polarization with increasing temperature reflects the thermal weakening of Zeeman-biased spin selectivity under thermodynamically consistent (i.e., consistent with the detailed balance) dissipative dynamics. Microscopically, polarization in this model is generated during charge transfer: interface-local spin-flip jumps convert singlet-born $DB$ population into triplet weight, and the local splitting $\Delta_1\propto B_{\rm eff}$ biases the relative likelihood of producing the $m=+1$ versus $m=-1$ triplet components. As $T$ increases ($\beta$ decreases), the detailed-balance asymmetry controlling this bias is progressively reduced, so the cycle generates a smaller net imbalance between the triplet projections per passage through the interface. Downstream, spin-conserving recombination returns whatever singlet and triplet character is present in the $DB$ manifold back into the corresponding donor channels, and donor-side relaxation closes the loop by converting triplet population back toward the singlet. Thus, when the interface bias is thermally washed out, these closure processes dominate, and the steady-state polarization drops toward a lower saturation value. Conversely, increasing $\alpha_0$ enhances $B_{\rm eff}$ and the associated interfacial splitting, strengthening the detailed-balance bias in the spin-conversion step and thereby sustaining a larger polarization at any given temperature. Taken together, the trends in Fig.~\ref{fig:pol_vs_temp} provide a direct, experimentally testable prediction for trESR/trEPR probes of the CISS cycle: raising $T$ suppresses the interface-generated triplet selectivity, while increasing the current-induced bias scale $\alpha_0$ counteracts that suppression.

{
}

\section{Conclusions and discussions}

In summary, we propose a spinterface mechanism operating entirely within a D--$\chi B$--A complex that can account for the magnitude and qualitative trends of the spin polarization reported in recent trEPR experiments, without invoking a phenomenological CISS parameter or requiring large intrinsic spin-orbit coupling on the bridge. Contrary to previous claims \cite{subotnik2023chiral}, such a spinterface model provides a mechanistic and first-principles understanding of emergent spin polarization in D--$\chi B$--A complexes with testable predictions.

In our two-electron Lindblad description, the donor electron acts as a localized moment that exchanges with an itinerant electron in the donor-bridge charge-transfer manifold, while the charge flow through the $\chi B$ generates an effective solenoidal field that sets the interfacial Zeeman bias $B_{\rm eff}$. Together with (i) interface-local, thermodynamically consistent spin-conversion at the donor-bridge interface and (ii) spin-conserving recombination followed by donor-side thermalization that closes the photochemical cycle, this bias is sufficient to produce tens-of-percent polarization over the 25-105~K window for a realistic range of parameters. 

By explicitly resolving the dependence on the solenoidal coupling $\alpha_0$, temperature $T$, and the mixing scale $\gamma_m$, the model identifies an optimal conversion regime in which the spin-mixing time $\tau_{\rm mix}=\hbar/\gamma_m$ is neither too long to generate appreciable triplet weight within a cycle nor too short to over-mix and depolarize the charge-transfer manifold; quantitatively, robust polarization emerges around $\gamma_m^\ast/t_{\rm hop}\sim 10^{-2}-10^{-1}$, i.e., $\tau_{\rm mix}^\ast\sim 10-10^2$ hopping times $\tau_{\rm hop}=\hbar/t_{\rm hop}$. Importantly, this sub-ps conversion window reflects fast interface/bridge electronic dynamics and should not be conflated with the experimentally reported bridge$\to$acceptor charge-transfer time, which is an environment-dressed, rate-limited kinetic scale. 

The coarse-grained dissipators used here naturally admit microscopic origins in transverse magnetic noise near the donor-bridge junction (e.g., hyperfine fields, $g$-strain/$\Delta g$ anisotropy, vibronic modulation of SOC, or paramagnetic impurities), and the present framework makes transparent how thermally weakened detailed-balance bias suppresses selectivity with increasing $T$, while increasing $\alpha_0$ counteracts that suppression by enhancing the interfacial splitting. At the same time, the minimal nature of our description (no explicit vibronic structure or molecular SOC on the bridge, a compact donor manifold, and phenomenological but thermodynamically consistent dissipation) highlights clear extensions: incorporating explicit vibronic/SOC pathways, treating multiple charge-transfer channels on equal footing, and extending the analysis to oriented ensembles where current direction and molecular handedness can be disentangled. More broadly, these results suggest {that the apparent observation of the CISS effect in D--$\chi B$--A complexes is supported by the spinterface mechanism as the origin of the CISS effect \cite{subotnik2023chiral, eckvahl_detecting_2024}. }
Further, donor-bridge internal ``spinterfaces'' may provide a unifying language for CISS in both device-based and molecule-internal platforms. 

Future experiments that systematically vary bridge-mediated charge-transfer times, donor-bridge coupling, or spin-mixing rates (e.g., via isotopic substitution or controlled changes of the chemical environment, tuning $\gamma_m$), all within the same D--$\chi$B--A platforms, could therefore test the internal-spinterface scenario by checking whether the apparent CISS fraction follows the trends with $\alpha_0$,  temperature, and $\gamma_m$ predicted here.

\section*{Methods}

All results are steady-state ($\dot\rho=0$) solutions of the Lindblad master equation of Sec.~\ref{sec:system}. The reduced two-electron basis comprises the doubly-occupied donor (DD) manifold and the site-resolved donor--bridge (DB) manifolds. The donor carries two orbitals $\alpha\in\{a,b\}$, so the two donor electrons span the full two-electron space of these orbitals, $\binom{4}{2}=6$ states (the singlet and three covalent triplets of
Sec.~\ref{sec:system}, plus the two doubly-occupied configurations $\lvert a\!\uparrow a\!\downarrow\rangle$, $\lvert b\!\uparrow b\!\downarrow\rangle$). Each DB manifold places one electron on the donor (orbital $a$ or $b$) and one on the bridge site $j$, giving $2\times(1\ \text{singlet}+3\ \text{triplets})=8$ states per site. 

The two doubly-occupied (ionic) donor configurations are excluded from the basis: pushed to high energy by the on-site Coulomb repulsion $U$, they are not connected to the remaining states by any term of $H$ or by any jump operator, and so remain unpopulated throughout the cycle.
Retaining only the dynamically coupled manifolds—the singlet and three triplets of the doubly-occupied donor and the site-resolved donor--bridge states—gives $d=4+8N=36$.

We solve for the steady state by vectorization. Using column-stacking, $\mathrm{vec}(\rho)$, the identity $\mathrm{vec}(AXB)=(B^{\mathsf T}\otimes A)\,\mathrm{vec}(X)$ maps the Lindbladian onto the superoperator
\begin{equation}
\mathcal{L}=-i\bigl(I\otimes H-H^{\mathsf T}\otimes I\bigr)
+\sum_{k}\Bigl[L_{k}^{*}\otimes L_{k}
-\tfrac{1}{2}\,I\otimes L_{k}^{\dagger}L_{k}
-\tfrac{1}{2}\,(L_{k}^{\dagger}L_{k})^{\mathsf T}\otimes I\Bigr],
\end{equation}
with $H$ and $\{L_{k}\}$ as defined in Sec.~\ref{sec:system}. For $d=36$ the vectorized state has length $d^{2}=1296$ and $\mathcal{L}$ is a $1296\times 1296$ matrix. The steady state is the (unique, up to normalization) null vector, $\mathcal{L}\,\mathrm{vec}(\rho_{SS})=0$, computed as an orthonormal basis of $\ker\mathcal{L}$ with an SVD-based \texttt{null} function; the resulting single basis column is reshaped to $36\times36$, Hermitianized, and normalized to $\mathrm{Tr}\,\rho_{SS}=1$.

Because the interfacial Zeeman term $H_{Z}$ depends on the steady-state bond currents $J_{\sigma}=\mathrm{Tr}[\hat{J}_{\sigma}\rho_{SS}]$, $\mathcal{L}$ is a functional of $\rho_{SS}$ and the null-space solve is embedded in a self-consistency loop: from a trial field ($J=0$) we build $\mathcal{L}$, solve for $\rho_{SS}$, re-evaluate $J_{\uparrow}+J_{\downarrow}$, update $H_{Z}$, and rebuild $\mathcal{L}$, iterating until $B_{eff}\propto\alpha_{0}J$ converges to a relative tolerance of $[10^{-12}]$ in the units of current. Observables ($\langle n_{N}\rangle$, $\langle S^{z}_{N}\rangle$, and the polarization $P$) are evaluated on the converged $\rho_{SS}$. Unless noted, $\gamma=\eta=\gamma_{0}=t_{hop}$ with $t_{hop}\approx80$~meV; $\alpha_{0}$, $\gamma_{m}$, and $T$ are varied as indicated per figure.

\section*{Associated Content}
\noindent\textbf{Supporting Information:} 
\noindent classical Biot--Savart estimation of the geometric solenoidal coupling parameter $\alpha_{\mathrm{geom}}$ for the PXX--NMI$_2$--NDI (PNN) triad; 
analysis of the non-linear torsional vibrational modulation of the spinterface field; 
and a discussion of the negligible intrinsic spin--orbit coupling on the NMI$_2$ bridge.

\section*{Acknowledgments}
OLAM gratefully acknowledges support from the U.S. National Science Foundation under grant no. CHE-2513261. Y.D. acknowledges support from a BSF grant No. 2023787

\bibliography{main, ref}

\newpage

\begin{figure*}
    \centering
    \includegraphics[width=3.25in, height=1.85in]
    {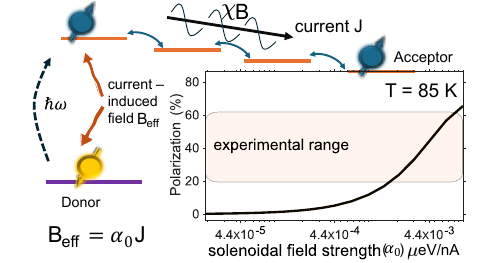}
    \caption{TOC Graphic}
\end{figure*}
\end{document}